# Constructions of Quasi-Twisted Two-Weight Codes

Eric Z. Chen
Dept. of Computer Science
Kristianstad University College
291 88 Kristianstad
Sweden
Email: eric.chen@hkr.se

*Abstract*— A code is said to be two-weight if the non-zero codewords have only two different a weight $w_1$ and $w_2$. Two-weight codes are closely related to strongly regular graphs. In this paper. It is shown that a consta-cyclic code of composite length can be put in the quasi-twisted form. Based on this transformation, a new construction method of quasi-twisted (QT) two-weight codes is presented. A large amount of QT two-weight codes are found, and some new codes are also constructed.

## I. INTRODUCTION

A linear *[n, k, d]$_q$* code [1] is a *k*-dimensional subspace of an *n*-dimensional vector space over *GF(q)*, with minimum distance *d* between any two codewords. As a generalization to cyclic codes, quasi-cyclic (QC) codes and quasi-twisted (QT) codes have been shown to contain many good linear codes. Many researchers have been using modern computers to search for good QC or QT codes, and many record-breaking codes are found.

A code is said to be two-weight if any non-zero codeword has a weight of $w_1$ or $w_2$. Two-weight codes are closely related to strongly regular graphs [6]. This paper is about the construction of two-weight codes.

The rest of the paper is organized as follows. Section II presents the introduction to consta-cyclic and quasi-twisted codes. In Section III, it is shown that a consta-cyclic code of composite length can be put into a quasi-cyclic form. In Section IV, a new construction method of quasi-twisted (QT) two-weight codes from consta-cyclic simplex codes is presented, and a large amount of QT two-weight codes are obtained. Some new codes are also constructed.

## II. CONSTA-CYCLIC CODES AND QUASI-TWISTED *CODES*

### A. *Cyclic Codes and Consta-Cyclic Codes*

A code is said to be cyclic if every cyclic shift of a codeword is also a codeword. A cyclic code can be described by the polynomial algebra. A cyclic *[n, k, d]$_q$* code has a unique generator polynomial *g(x)*. It is a polynomial with degree of *n – k*. All codewords of a cyclic code are multiples of g(x) modulo $x^n – 1$. A linear *[n, k, d]$_q$* code is said to be *λ*-consta-cyclic if there is a non-zero element *λ* of GF(q) such that for any codeword ($a_0$, $a_1$, ..., $a_{n-1}$), a consta-cyclic shift by one position or ($\lambda a_{n-1}$, $a_0$, ..., $a_{n-2}$) is also a codeword [2]. Therefore, the consta-cyclic code is a generalization of the cyclic code, or a cyclic code is a *λ*-consta-cyclic code with *λ* = 1. Similarly, a consta-cyclic code can be defined by a generator polynomial.

### B. *Quasi-Twisted Codes*

A code is said to be quasi-twisted (QT) if a consta-cyclic shift of any codeword by *p* positions is still a codeword. Thus a consta-cyclic code is a QT code with *p = 1*, and a quasi-cyclic (QC) code is a QT code with *λ = 1*. The length *n* of a QT code is a multiple of *p*, i.e., *n = mp*.

The consta-cyclic matrices are also called twistulant matrices. They are basic components in the generator matrix for a QT code. An m × m consta-cyclic matrix is defined as

$$C = \begin{bmatrix} c_0 & c_1 & c_2 & \cdots & c_{m-1} \\ \lambda c_{m-1} & c_0 & c_1 & \cdots & c_{m-2} \\ \lambda c_{m-2} & \lambda c_{m-1} & c_0 & \cdots & c_{m-3} \\ \vdots & \vdots & \vdots & \vdots & \vdots \\ \lambda c_1 & \lambda c_2 & \lambda c_3 & \cdots & c_0 \end{bmatrix}, \quad (1)$$

and the algebra of m × m consta-cyclic matrices over *GF(q)* is isomorphic to the algebra in the ring f[x]/($x^m$–λ) if C is mapped onto the polynomial formed by the elements of its first row, $c(x) = c_0 + c_1 x + \ldots + c_{m-1} x^{m-1}$, with the least significant coefficient on the left. The polynomial $c(x)$ is also called the defining polynomial of the matrix C. A twistulant matrix is called a circulant matrix if $\lambda = 1$.

The generator matrix of a QT code can be transformed into rows of twistulant matrices by suitable permutation of columns [3]. For examples, A 1-generator QT $[m\ p,\ k]_q$ code has the following form of the generator matrix [4]:

$$G = [\ G_0\ G_1\ G_2\ \ldots\ G_{p-1}\ ] \quad (2)$$

where $G_i$, $i = 0, 1, 2, \ldots, p-1$, are circulant matrices of order m. Let $g_0(x), g_1(x), \ldots, g_{p-1}(x)$ are the corresponding defining polynomials. A 2-generator QT $[m\ p,\ k]_q$ codes has the generator matrix of the following form:

$$G = \begin{bmatrix} G_{00} & G_{01} & \ldots & G_{0,p-1} \\ G_{10} & G_{11} & \ldots & G_{1,p-1} \end{bmatrix}, \quad (3)$$

where $G_{ij}$ are circular matrices, for $i = 0$, and $1$, $j = 0, 1, \ldots, p-1$.

### III. QUASI-TWISTED STRUCTURE OF A CONSTA-CYCLIC CODE

*A. Consta-Twisted Form of a consta-cyclic Matrix*

In [5], it was shown that a circulant matrix can be put into a quasi-cyclic (QC) form. It is obvious that a direct generalization can be made to transform a consta-cyclic matrix into a quasi-twisted (QT) form.

Consider an n×n consta-cyclic matrix C with n = mr. Let $c(x)$ be its defining polynomial. We number the rows and columns with 0, 1, 2, ..., mr – 1. To obtain a quasi-twisted structure form of such a consta-cyclic matrix, we do the row and column permutations. If we re-order the rows and columns in the following order: 0, r, ..., (m-1)r, 1, r+1, (m-1)r+1, ..., (r-1), r +(r-1), ..., (m-1)r + (r-1). Then we obtain a matrix that consists of $r$ rows of $r$ twistulant matrices of order m. If we use the defining polynomials to represent the first row of the twistulant matrices, the QT form of the consta-cyclic matrix can be represented by these defining polynomials. To make it clear, we use an example.

**Example** Let n = 21 = mr = 3 × 7, m = 3, r = 7, and q = 4. Let $0, 1, \alpha$, and $\beta = 1 + \alpha$ be elements of $GF(4)$ and $\lambda = \beta$. Then the $\lambda$-consta-cyclic matrix defined by $c(x) = 1 + \beta x + \beta x^3 + \beta x^4 + \beta x^5 + \alpha x^6 + x^7 + x^8 + \alpha x^9 + x^{10} + \alpha x^{11} + x^{13} + \alpha x^{15} + \beta x^{16} + x^{17} + x^{18}$ is given as follows. In fact, this polynomial is also a generator polynomial for the consta-cyclic $[21, 3, 16]_4$ code, and the following matrix is its generator matrix.

$$C = \begin{bmatrix}
1 & \beta & 0 & \beta & \beta & \beta & \alpha & 1 & 1 & \alpha & 1 & \alpha & 0 & 1 & 0 & \alpha & \beta & 1 & 1 & 0 & 0 \\
0 & 1 & \beta & 0 & \beta & \beta & \beta & \alpha & 1 & 1 & \alpha & 1 & \alpha & 0 & 1 & 0 & \alpha & \beta & 1 & 1 & 0 \\
0 & 0 & 1 & \beta & 0 & \beta & \beta & \beta & \alpha & 1 & 1 & \alpha & 1 & \alpha & 0 & 1 & 0 & \alpha & \beta & 1 & 1 \\
\beta & 0 & 0 & 1 & \beta & 0 & \beta & \beta & \beta & \alpha & 1 & 1 & \alpha & 1 & \alpha & 0 & 1 & 0 & \alpha & \beta & 1 \\
\beta & \beta & 0 & 0 & 1 & \beta & 0 & \beta & \beta & \beta & \alpha & 1 & 1 & \alpha & 1 & \alpha & 0 & 1 & 0 & \alpha & \beta \\
\alpha & \beta & \beta & 0 & 0 & 1 & \beta & 0 & \beta & \beta & \beta & \alpha & 1 & 1 & \alpha & 1 & \alpha & 0 & 1 & 0 & \alpha \\
1 & \alpha & \beta & \beta & 0 & 0 & 1 & \beta & 0 & \beta & \beta & \beta & \alpha & 1 & 1 & \alpha & 1 & \alpha & 0 & 1 & 0 \\
0 & 1 & \alpha & \beta & \beta & 0 & 0 & 1 & \beta & 0 & \beta & \beta & \beta & \alpha & 1 & 1 & \alpha & 1 & \alpha & 0 & 1 \\
\beta & 0 & 1 & \alpha & \beta & \beta & 0 & 0 & 1 & \beta & 0 & \beta & \beta & \beta & \alpha & 1 & 1 & \alpha & 1 & \alpha & 0 \\
0 & \beta & 0 & 1 & \alpha & \beta & \beta & 0 & 0 & 1 & \beta & 0 & \beta & \beta & \beta & \alpha & 1 & 1 & \alpha & 1 & \alpha \\
1 & 0 & \beta & 0 & 1 & \alpha & \beta & \beta & 0 & 0 & 1 & \beta & 0 & \beta & \beta & \beta & \alpha & 1 & 1 & \alpha & 1 \\
\beta & 1 & 0 & \beta & 0 & 1 & \alpha & \beta & \beta & 0 & 0 & 1 & \beta & 0 & \beta & \beta & \beta & \alpha & 1 & 1 & \alpha \\
1 & \beta & 1 & 0 & \beta & 0 & 1 & \alpha & \beta & \beta & 0 & 0 & 1 & \beta & 0 & \beta & \beta & \beta & \alpha & 1 & 1 \\
\beta & 1 & \beta & 1 & 0 & \beta & 0 & 1 & \alpha & \beta & \beta & 0 & 0 & 1 & \beta & 0 & \beta & \beta & \beta & \alpha & 1 \\
\beta & \beta & 1 & \beta & 1 & 0 & \beta & 0 & 1 & \alpha & \beta & \beta & 0 & 0 & 1 & \beta & 0 & \beta & \beta & \beta & \alpha \\
1 & \beta & \beta & 1 & \beta & 1 & 0 & \beta & 0 & 1 & \alpha & \beta & \beta & 0 & 0 & 1 & \beta & 0 & \beta & \beta & \beta \\
\alpha & 1 & \beta & \beta & 1 & \beta & 1 & 0 & \beta & 0 & 1 & \alpha & \beta & \beta & 0 & 0 & 1 & \beta & 0 & \beta & \beta \\
\alpha & \alpha & 1 & \beta & \beta & 1 & \beta & 1 & 0 & \beta & 0 & 1 & \alpha & \beta & \beta & 0 & 0 & 1 & \beta & 0 & \beta \\
\alpha & \alpha & \alpha & 1 & \beta & \beta & 1 & \beta & 1 & 0 & \beta & 0 & 1 & \alpha & \beta & \beta & 0 & 0 & 1 & \beta & 0 \\
0 & \alpha & \alpha & \alpha & 1 & \beta & \beta & 1 & \beta & 1 & 0 & \beta & 0 & 1 & \alpha & \beta & \beta & 0 & 0 & 1 & \beta \\
\alpha & 0 & \alpha & \alpha & \alpha & 1 & \beta & \beta & 1 & \beta & 1 & 0 & \beta & 0 & 1 & \alpha & \beta & \beta & 0 & 0 & 1
\end{bmatrix}$$

After re-ordering of rows and columns, we get the following matrix consisting of 7 rows of twistulant matrices of order 3:

$$A = \begin{bmatrix}
1 & 1 & 0 & \beta & 1 & \alpha & 0 & \alpha & \beta & \beta & 1 & 1 & \beta & \alpha & 1 & \beta & 0 & 0 & \alpha & 1 & 0 \\
0 & 1 & 1 & 1 & \beta & 1 & \alpha & 0 & \alpha & \beta & \beta & 1 & \beta & \beta & \alpha & 0 & \beta & 0 & 0 & \alpha & 1 \\
\beta & 0 & 1 & \beta & 1 & \beta & 1 & \alpha & 0 & \beta & \beta & \beta & 1 & \beta & \beta & 0 & 0 & \beta & \beta & 0 & \alpha \\
0 & \alpha & 1 & 1 & 1 & 0 & \beta & 1 & \alpha & 0 & \alpha & \beta & \beta & 1 & 1 & \beta & \alpha & 1 & \beta & 0 & 0 \\
\beta & 0 & \alpha & 0 & 1 & 1 & 1 & \beta & 1 & \alpha & 0 & \alpha & \beta & \beta & 1 & \beta & \beta & \alpha & 0 & \beta & 0 \\
1 & \beta & 0 & \beta & 0 & 1 & \beta & 1 & \beta & 1 & \alpha & 0 & \beta & \beta & \beta & 1 & \beta & 0 & 0 & 0 & \beta \\
0 & \beta & 0 & 0 & \alpha & 1 & 1 & 1 & 0 & \beta & 1 & \alpha & 0 & \alpha & \beta & \beta & 1 & 1 & \beta & \alpha & 1 \\
0 & 0 & \beta & \beta & 0 & \alpha & 0 & 1 & 1 & 1 & \beta & 1 & \alpha & 0 & \alpha & \beta & \beta & 1 & \beta & \beta & \alpha \\
\alpha & 0 & 0 & 1 & \beta & 0 & \beta & 0 & 1 & \beta & 1 & \beta & 1 & \alpha & 0 & \beta & \beta & \beta & 1 & \beta & \beta \\
\beta & \beta & \alpha & 0 & \beta & 0 & 0 & \alpha & 1 & 1 & 1 & 0 & \beta & 1 & \alpha & 0 & \alpha & \beta & \beta & 1 & 1 \\
1 & \beta & \beta & 0 & 0 & \beta & \beta & 0 & \alpha & 0 & 1 & 1 & 1 & \beta & 1 & \alpha & 0 & \alpha & \beta & \beta & 1 \\
\alpha & 1 & \beta & \alpha & 0 & 0 & 1 & \beta & 0 & \beta & 0 & 1 & \beta & 1 & \beta & 1 & \alpha & 0 & \beta & \beta & \beta \\
\beta & \beta & 1 & \beta & \beta & \alpha & 0 & \beta & 0 & 0 & \alpha & 1 & 1 & 1 & 0 & \beta & 1 & \alpha & 0 & \alpha & \beta \\
\beta & \beta & \beta & 1 & \beta & \beta & 0 & 0 & \beta & \beta & 0 & \alpha & 0 & 1 & 1 & 1 & \beta & 1 & \alpha & 0 & \alpha \\
\alpha & \beta & \beta & \alpha & 1 & \beta & \beta & 0 & 0 & 1 & \beta & 0 & \beta & 0 & 1 & \beta & 1 & \beta & 1 & \alpha & 0 \\
\alpha & 0 & \alpha & \beta & \beta & 1 & \beta & \beta & \alpha & 0 & \beta & 0 & 0 & \alpha & 1 & 1 & 1 & 0 & \beta & 1 & \alpha \\
1 & \alpha & 0 & \beta & \beta & \beta & 1 & \beta & \beta & 0 & 0 & \beta & \beta & 0 & \alpha & 0 & 1 & 1 & 1 & \beta & 1 \\
0 & 1 & \alpha & \alpha & \beta & \beta & \alpha & 1 & \beta & \alpha & 0 & 0 & 1 & \beta & 0 & \beta & 0 & 1 & \beta & 1 & \beta \\
1 & \beta & 1 & \alpha & 0 & \alpha & \beta & \beta & 1 & \beta & \beta & \alpha & 0 & \beta & 0 & 0 & \alpha & 1 & 1 & 1 & 0 \\
\beta & 1 & \beta & 1 & \alpha & 0 & \beta & \beta & \beta & 1 & \beta & \beta & 0 & 0 & \beta & \beta & 0 & \alpha & 0 & 1 & 1 \\
\alpha & \beta & 1 & 0 & 1 & \alpha & \alpha & \beta & \beta & \alpha & 1 & \beta & \alpha & 0 & 0 & 1 & \beta & 0 & \beta & 0 & 1
\end{bmatrix}$$

So, for i = 1, 2, ..., r = 7, if we use $a_i(x)$ to denote the defining polynomials for the first row of twistulant matrices, we have $a_1(x) = 1 + x$, $a_2(x) = \beta + \alpha x + x^2$, $a_3(x) = \alpha x + \beta x^2$, $a_4(x) = \beta + x + x^2$, $a_5(x) = \beta + \alpha x + x^2$, $a_6(x) = \beta$, and $a_7(x) = \alpha + x$. The matrix A can be specified by the following matrix of defining polynomials. So the matrix A can be specified by r defining polynomials.

$$A(x) = \begin{bmatrix} a_1(x) & a_2(x) & a_3(x) & \ldots & a_r(x) \\ xa_r(x) & a_1(x) & a_2(x) & \ldots & a_{r-1}(x) \\ xa_{r-2}(x) & xa_{r-1}(x) & a_1(x) & \ldots & a_{r-2}(x) \\ \vdots & \vdots & \vdots & \vdots & \vdots \\ xa_2(x) & xa_3(x) & xa_4(x) & \ldots & a_1(x) \end{bmatrix},$$

(4)

where the computation is done with modulo $x^m - \lambda$.

### B. Quasi-Twisted Structure of Simplex Codes

It is well known that for any positive integer $t > 1$ and prime power q, we have a Hamming $[n, n-t, 3]_q$ code, where $n = (q^t-1)/(q-1)$. Further, if t and $q-1$ are relatively prime, then the Hamming code is equivalent to a cyclic code. The dual code of a Hamming code is called the simplex code. So for any integer $t > 1$ and prime power q, there is a simplex $[(q^t-1)/(q-1), t, q^{t-1}]_q$ code.

A simplex code can be constructed as a consta-cyclic code [2]. Let h(x) be a primitive polynomial of degree $t$ over GF(q). A $\lambda$-consta-cyclic simplex $[(q^t-1)/(q-1), t, q^{t-1}]_q$ code can be defined by the generator polynomial g(x) = $(x^n-\lambda)/h(x)$, where $n = (q^t-1)/(q-1)$, and $\lambda$ is a non-zero element of GF(q) and has order of $q-1$. Further, a $\lambda$-consta-cyclic simplex code is cyclic if t and $q-1$ are relatively prime.

It should be noted that a simplex code is an equidistance code, where $q^t-1$ non-zero codewords have a weight of $q^{t-1}$. The $q^t-1$ non-zero codewords are rows given by the consta-cyclic matrix defined by the generator polynomial, and their multiples by non-zero elements of GF(q).

If $n = (q^t-1)/(q-1)$ is not prime, we write $n = mr$. let g(x) be a generator polynomial of such a $\lambda$-consta-cyclic simplex $[(q^t-1)/(q-1), t, q^{t-1}]_q$ code and G be the consta-cyclic matrix defined by g(x). Then using the method given, G can be put into quasi-twisted form. It consists of $r \times r$ matrix of consta-cyclic matrices of order m. Therefore, the $\lambda$-consta-cyclic simplex $[(q^t-1)/(q-1), t, q^{t-1}]_q$ code is equivalent to a QT $[mr, t, q^{t-1}]_q$ code. In the above example, c(x) is the generator polynomial for a consta-cyclic $[21, 3, 16]_4$ code. For $m = 3$, we obtain the equivalent QT simplex $[3 \times 7, 3, 16]_4$ code, after the column permutation. It can be shown that it is 1-generator QC code with defining polynomials $a_i(x)$ as given in the example, $i = 1, 2, ..., r$.

In practice, given the generator polynomial of a consta-cyclic simplex code of composite length, we can obtain the defining polynomials $a_i(x)$ directly, and then A(x) too.

## IV. CONSTRUCTIONS OF QT TWO-WEIGHT CODES

### A. Two-Weight Codes

A linear code is called projective if any two of its coordinates are linearly independent, or in other words, if the minimum distance of its dual code is at least three. A code is said to be two-weight if any non-zero codeword has a weight of $w_1$ or $w_2$, where $w_1 \neq w_2$. A two weight code can also be written as the $[n, k; w_1, w_2]_q$ code. Two-weight codes are closely related to strongly regular graphs.

In the survey paper [6], Calderbank and Kantor presented many known families of two-weight codes. An online database of two-weight codes is also available [7]. In this section, a new construction method is presented based on quasi-twisted structure of consta-cyclic simplex codes of composite length, and a large amount of QT two-weight codes are found and listed.

### B. Construction of QT Two-Weight Codes

In [8], a special family of two-weight codes, named SU2, was studied and an explicit construction of the codes was given. They are 2-generator quasi-twisted codes. Quasi-cyclic two-weight codes can also be constructed from irreducible cyclic codes [9]. Gulliver [10, 11] presented 3 new quasi-cyclic two-weight codes. Recently, Kohnert [12] has presented a construction of two-weight codes with prescribed groups of automorphisms. So it is interesting to turn attention to codes with only two non-zero weights without focusing on the minimum distance. This paper is the attempt in this direction.

Let g(x) be the generator polynomial of a consta-cyclic simplex $[(q^k-1)/(q-1), k, q^{k-1}]_q$ code of composite length, and G be the consta-cyclic matrix defined by g(x). Let $n = (q^k-1)/(q-1) = mr$. Using the method given in the last section, we can put G into r rows of r twistulant matrices. Let $a_i(x)$ be r defining polynomials for the first row of the twistulant matrices, $i = 1, 2, ..., r$. Then we get a matrix of defining polynomials as given in (4). Let $d_i$ be the weights of the defining polynomial $a_i(x)$, $i = 1, 2, ..., r$. We denote W the weight matrix corresponding to the weights of the defining polynomials of A(x). Based on the structure of A(x), the i-th row of W is the cyclic shift by one position of the $(i-1)$-th row, $i = 2, 3, ..., r$.

**Example** (continued) By computing the weights of defining polynomials in the first row of A(x), we obtain the weight matrix W for A(x). It is a cyclic matrix.

$$W = \begin{bmatrix} 2 & 3 & 2 & 3 & 3 & 1 & 2 \\ 2 & 2 & 3 & 2 & 3 & 3 & 1 \\ 1 & 2 & 2 & 3 & 2 & 3 & 3 \\ 3 & 1 & 2 & 2 & 3 & 2 & 3 \\ 3 & 3 & 1 & 2 & 2 & 3 & 2 \\ 2 & 3 & 3 & 1 & 2 & 2 & 3 \\ 3 & 2 & 3 & 3 & 1 & 2 & 2 \end{bmatrix}, \quad (5)$$

For the given weight matrix W, to search for a QT two-weight *[mp, k, d]$_q$* code, we need to find *p* columns of W, such that the row sums of the selected columns can only be of two non-zero values ($w_1$ or $w_2$).

If columns $c_1, c_2, …, c_p$ produce the row sums of two values, then the complement columns (the unselected columns) produce also the row sums of two values, since the sum of any row of W has the same value of d = $q^{t-1}$. Therefore, among r columns, it is sufficient to search for QT two-weight *[mp, k, d]$_q$* codes with p ≤ $\lfloor r/2 \rfloor$, where $\lfloor x \rfloor$ denotes the largest integer less than or equal to x. When the columns are found, the corresponding columns in the defining polynomial matrix give the defining polynomials of the QT two-weight code. The procedure given in [5] can be used to decide if the code is a 1-generator, or 2-generator, or g-generator QT code.

**Example** (continued) The weight matrix from the cosnta-cyclic simplex [21, 3, 16]$_4$ code is given in (5). The columns 1, 2, 4 of W produce the row sums of 8, 6, 6, 6, 8, 6, 8, respectively. So these row sums are of values 6 and 8. Therefore, columns 1, 2, and 4 defines a QT two-weight [9, 3; 6, 8]$_4$ code. The complement columns are columns 3, 5, 6 and 7. The row sums are 8, 10, 10, 10, 8, 10, 8. That is, the row sums are of values 8 and 10. They define a QT two-weight [12, 3; 8, 10]$_4$ code. Their generator matrices are given by the corresponding columns in A, and can be found to be as follows (by using the method given in [5]):

$$G = (a_1(x); a_2(x), a_4(x)) = \begin{bmatrix} 1 & 1 & 0 & \beta & 1 & \alpha & \beta & 1 & 1 \\ 0 & 1 & 1 & 1 & \beta & 1 & \beta & \beta & 1 \\ \beta & 0 & 1 & \beta & 1 & \beta & \beta & \beta & \beta \end{bmatrix},$$

$$G = (a_3(x); a_5(x); a_6(x); a_7(x))$$
$$= \begin{bmatrix} 0 & \alpha & \beta & \beta & \alpha & 1 & \beta & 0 & 0 & \alpha & 1 & 0 \\ \alpha & 0 & \alpha & \beta & \beta & \alpha & 0 & \beta & 0 & 0 & \alpha & 1 \\ 1 & \alpha & 0 & 1 & \beta & \beta & 0 & 0 & \beta & \beta & 0 & \alpha \end{bmatrix},$$

### C. Computation Results on Two-Weight Codes

Since the weight matrix is cyclic and only r elements of it are needed in memory. Other columns can be obtained from these r elements. So space is not a problem. The problem is time. For small r and reasonable size, an exhaustive search is used. For large r, the exhaustive is not possible, so a limited search is used. Extensive computer search has been made and a large amount of quasi-twisted two-weight codes have been found. Although many codes found have the same parameters as known ones, most of them are constructed in the quasi.-twisted form. Due to space limit, only selected results are given in Tables I, II. More results can be obtained from the author. All SU2 family of two-weight codes is found, and they are not listed in the tables. The original sources of known codes can be found by looking them up in the online database [7].

## V. CONCLUSION

In this paper, codes with only two weights are studied. Based on quasi-twisted form of a consta-cyclic simplex code, a computer construction of quasi-twisted two-weight codes is presented. A detailed example is given to show the method. A lot of quasi-twisted two-weight codes have been found, and several new codes are also found. More research in the method is deserved.

TABLE I
Binary QC Two-Weight *[mp, k]* CODES

| k | m | p  | $w_1, w_2$ | Note |
|---|---|----|------------|------|
| 4 | 5 | 1  | 2, 4       | [6]  |
|   |   | 2  | 4, 6       | [6]  |
| 6 | 3 | 6  | 8, 12      | [6]  |
|   |   | 7  | 8, 12      | [6]  |
|   |   | 9  | 12, 16     | [6]  |
|   |   | 12 | 16, 20     | [6]  |
|   |   | 14 | 20, 24     | [6]  |
|   |   | 15 | 20, 24     | [6]  |
|   |   | 16 | 24, 32     | [6]  |
|   |   | 20 | 30, 32     | [6]  |
| 8 | 17| 3  | 24, 32     | [6]  |
|   |   | 4  | 32, 40     | [6]  |
|   |   | 5  | 40, 48     | [6]  |

TABLE I
q-ary QT Two-Weight *[mp, k]* CODES

| q | k | m | p | w₁, w₂ | Note |
|---|---|---|---|---|---|
| 3 | 4 | 4 | 2 | 3, 6 | [6] |
|   |   | 5 | 2 | 6, 9 | [6] |
|   |   |   | 3 | 9, 12 | [6] |
|   |   |   | 4 | 12, 15 | [6] |
|   |   |   | 5 | 15, 18 | [6] |
|   |   |   | 6 | 18, 21 | [6] |
|   | 5 | 11 | 1 | 6, 9 | [6] |
|   | 5 | 11 | 5 | 36, 45 | [7] |
|   | 5 | 11 | 6 | 36, 45 | [7] |
|   | 5 | 11 | 10 | 72, 75 | [6] |
|   | 6 | 7 | 8 | 36, 45 | [6] |
|   |   |   | 12 | 54, 63 | [10] |
|   |   |   | 13 | 54, 63 | [7] |
|   |   |   | 14 | 63, 72 | [10] |
|   |   |   | 16 | 72, 81 | [7] |
|   |   |   | 18 | 81, 90 | [7] |
|   |   |   | 20 | 90, 99 | [7] |
|   |   |   | 22 | 99, 108 | [11] |
|   |   |   | 24 | 108, 117 | [7] |
|   |   |   | 26 | 117, 126 | [7] |
|   |   |   | 28 | 126, 135 | [7] |
|   |   |   | 30 | 135, 144 | [7] |
|   |   |   | 32 | 144, 153 | [7] |
|   |   |   | 6 | 48, 56 | [6] |
|   |   |   | 7 | 56, 64 | [6] |
|   |   |   | 8 | 64, 72 | [6] |
|   |   |   | 9 | 72, 80 | [6] |
|   |   |   | 10 | 80, 88 | [6] |
|   |   |   | 11 | 88, 96 | [6] |
|   |   |   | 12 | 96, 104 | [6] |
| 9 | 7 |   | 10 | 32, 40 | [7] |
|   |   |   | 63 | 216, 224 | [7] |
| 10 | 33 |   | 15 | 240, 256 | [6] |
|   |   |   | 16 | 256, 272 | [6] |
| 11 | 23 |   | 12 | 128, 144 | [6] |
|   |   |   | 77 | 880, 896 | [6] |
| 12 | 91 |   | 3 | 128, 144 | [7] |
|   |   |   | 5 | 224, 256 | [6] |
|   |   |   | 10 | 448, 480 | [6] |
|   |   |   | 15 | 672, 704 | [6] |
|   |   |   | 20 | 896, 928 | [6] |
|   |   |   | 42 | 1904, 1920 | [7] |
|   | 117 |   | 2 | 112, 128 | [7] |
|   |   |   | 10 | 576, 608 | [12] |
|   |   |   | 15 | 864, 996 | [12] |
|   |   |   | 33 | 1920, 1936 | [7] |
|   |   |   | 34 | 153, 162 | [7] |
|   |   |   | 36 | 162, 171 | [7] |
|   |   |   | 38 | 171, 180 | [7] |
|   |   |   | 39 | 180, 189 | [7] |
|   |   |   | 40 | 180, 189 | [7] |
|   |   |   | 44 | 198, 207 | [7] |
|   | 12 | 3796 | 2 | 5022, 5103 | new |
| 4 | 6 | 39 | 10 | 288, 304 | new |
|   |   |   | 15 | 432, 448 | new |
|   |   |   | 20 | 576, 592 | new |
|   |   |   | 25 | 720, 736 | new |
| 8 | 4 | 13 | 3 | 32, 36 | new |
|   |   |   | 42 | 476, 480 | new |
| 9 | 4 | 41 | 7 | 215, 225 | new |
|   |   |   | 13 | 504, 513 | new |
| 13 | 4 | 119 | 5 | 546, 559 | new |
|   |   |   | 15 | 1638, 1651 | new |